\documentclass{jpp}

\usepackage[utf8]{inputenc}
\usepackage[T1]{fontenc}
\usepackage{amsfonts,amsmath,amssymb}
\usepackage{bm,nicefrac}
\usepackage{graphicx}
\usepackage{tabularx}
\usepackage{subcaption}
\usepackage[section]{placeins}
\usepackage{textgreek}
\usepackage[breaklinks=true,colorlinks=true,linkcolor=blue,urlcolor=blue,citecolor=blue]{hyperref} 
\usepackage[table]{xcolor}
\usepackage[english]{babel}
\usepackage{listings}

\DeclareMathOperator*{\argmin}{arg\,min}

\definecolor{my-gray}{gray}{0.95}
\definecolor{codegreen}{rgb}{0,0.6,0}
\definecolor{codegray}{rgb}{0.5,0.5,0.5}
\definecolor{codepurple}{rgb}{0.58,0,0.82}
\definecolor{codered}{rgb}{0.95,0.02,0.03}

\lstdefinestyle{mystyle}{
    backgroundcolor=\color{my-gray},   
    commentstyle=\color{codegreen},
    keywordstyle=\color{codepurple},
    numberstyle=\tiny\color{codegray},
    stringstyle=\color{codered},
    basicstyle=\ttfamily\footnotesize,
    breakatwhitespace=false,         
    breaklines=true,                 
    captionpos=b,                    
    keepspaces=true,                 
    numbers=left,                    
    numbersep=5pt,                  
    showspaces=false,                
    showstringspaces=false,
    showtabs=false,                  
    tabsize=2
}

\title{The DESC Stellarator Code Suite Part II: Perturbation and continuation methods}

\author{Rory Conlin \aff{1}  \corresp{\email{wconlin@princeton.edu}},
Daniel W. Dudt \aff{1}, 
Dario Panici \aff{1},
\and Egemen Kolemen \aff{1} \corresp{\email{ekolemen@princeton.edu}}}

\affiliation{\aff{1}Princeton University, Princeton, New Jersey 08544}

\begin{document}

\maketitle

\begin{abstract}

A new perturbation and continuation method is presented for computing and analyzing stellarator equilibria. The method is formally derived from a series expansion about the equilibrium condition $\bm{F} \equiv \bm{J}\times \bm{B} - \nabla p = 0$, and an efficient algorithm for computing solutions to 2nd and 3rd order perturbations is developed. The method has been implemented in the DESC stellarator equilibrium code, using automatic differentiation to compute the required derivatives. Examples are shown demonstrating its use for computing complicated equilibria, perturbing a tokamak into a stellarator, and performing parameter scans in pressure, rotational transform and boundary shape in a fraction of the time required for a full solution.
\end{abstract}

\section{Introduction}
\label{sec:intro}

In the search for controlled nuclear fusion, 3D magnetic confinement devices such as stellarators have been shown to have several advantages over 2D magnetic geometries such as tokamaks, such as lower risk of disruption \citep{helander2012} and current-free steady state operation \citep{helander2014}. However, achieving good performance in a stellarator often requires significant optimization of the plasma equilibrium.
An additional advantage is that because of the generally lower plasma current and consequently fewer instabilities, more of the design and optimization can be done computationally, without requiring building and testing full scale devices \citep{boozer2015}. Despite this, designing and optimizing 3D magnetic equilibria that have good properties is still a computationally intensive task, for which a number of codes and software packages have been developed \citep{dudt2020desc,vmec1983,simsopt2021,spec2011,spong1998stellopt,  stellopt,rose2019}

Perturbation methods have been used heavily in tokamak plasma physics, primarily to analyze the stability of axisymmetric MHD equilibria by searching for a perturbation that minimizes the energy of the plasma, as in Bernstein's energy principle \citep{bernstein1958}. This has been extended to a wide range of of codes for analyzing fusion devices under small perturbations to an MHD equilibrium, such as the GPEC suite of codes \citep{glasser2016dcon,glasser2018stride,park2007,park2009} for analyzing tokamak configurations under 3D perturbations. A common feature of perturbation methods is using local approximation methods (commonly Taylor series) to examine solutions nearby to some equilibrium in an infinitesimal limit. In the present work, we extend this to include finite magnitude perturbations, and extend the approximation to 2nd and higher order to achieve increased accuracy for finite step sizes. While a perturbation usually refers to a single step in parameter space, a sequence of perturbations can be combined into a continuation method to further explore the phase space, where a single perturbation step is used to approximate a nearby solution, and Taylor approximation is recomputed at the new point before performing another finite size perturbation and so on.

Continuation methods have received less attention in the fusion community, though they have seen extensive use in other fields such as\citep{howell_computation_2009,richter1983continuation}. For the purposes of the present work, continuation methods can be used to solve parameterized equations of the form $F(x, \eta)=0$, where we identify $x$ as the solution vector, and $\eta$ as a continuation parameter. When $\eta$ is varied we find a family of solutions connected in parameter space, as well as possible branches and bifurcations of this family, indicated by points where the Jacobian of $F$ is singular. Starting from the initial value $(x_0, \eta_0)$ where $F$ is zero, we continuously vary $\eta$ while simultaneously varying $x$ such that the equation $F(x,\eta)=0$ is satisfied. Various methods exist for finding the solution curve (the locus of points $(x, \eta)$ where $F(x,\eta)=0$ is satisfied) such as piecewise linear (simplex) continuation and psuedo-arclength methods \citep{allgower_continuation_book_1990}. In an abstract sense, the standard multigrid method for solving partial differential equations can be though of as a discrete continuation method, where the continuation parameter $\eta$ governs the level of numerical resolution (ie grid spacing, number of basis functions etc.)

In general, existing codes for computing stellarator equilibria \citep{vmec1983,spec2011} only find discrete equilibria, and solving for a new equilibrium requires running the code from scratch (possibly with a different starting guess). Several questions that may then be asked are:
\begin{enumerate}
    \item Can these equilibria be computed more efficiently?\\
    \item Given a single equilibrium solution, can we find other similar solutions? \\
    \item What does the full phase space of 3D MHD equilibria look like? \\
\end{enumerate}

This is part II of a three-part series of papers on the DESC stellarator optimization code suite. Part I details the DESC equilibrium solver in comparison with the VMEC \citep{hirshman_steepestdescent_1983} 3D MHD equilibrium code. Computing 3D MHD equilibria is the preliminary step in stellarator analysis, so computing these equilibria quickly and accurately is important for further studies. In this paper, we describe a new continuation method for computing complicated stellarator equilibria using perturbations that attempts to resolve these questions. In \autoref{sec:code_improvements} we describe the DESC code and why it is the ideal code for implementation of the methods described in this paper, while in \autoref{sec:perturbations} we describe the mathematical background to the perturbation method. In \autoref{sec:continuation} demonstrate how to use these perturbations in a continuation method for computing stellarator equilibria. In \autoref{sec:applications} we demonstrate other applications of the perturbation method for computing and analyzing stellarator equilibria. Part III \citep{dudt2022desc} presents DESC's unique stellarator optimization capabilities made possible by the efficient equilibrium solver and the perturbation method, resulting in orders of magnitude speed-up in optimization. These advantages are shown in the context of quasi-symmetry optimization, where results are compared to conventional tools \citep{spong1998stellopt}. Three different quasi-symmetry objective formulations are also shown, with the relative advantages of each compared, highlighting the flexibility of DESC as an optimization code.

\section{The DESC code} \label{sec:code_improvements}

DESC is a recently developed \citep{dudt2020desc} pseudo-spectral code for computing 3D MHD equilibria. DESC computes 3D MHD equilibia by solving the force balance equations $\bm{J} \times \bm{B} - \nabla p = 0$ as opposed to the more common variational method which minimizes the MHD energy $\int_V (B^2/2\mu_0 + p) \ dV$. The independent variables in the equation are the positions of the flux surfaces $(R,Z)$ as well as the stream function $\lambda$, defined as the difference between the boundary poloidal angle $\theta$ and the straight field line poloidal angle: $\vartheta = \theta + \lambda$. These quantities are discretized in a Fourier-Zernike basis, using a Fourier series in the toroidal direction and Zernike polynomials in the poloidal/radial directions. After discretization, the equilibrium equation is expressed as a set of nonlinear algebraic equations $\bm{f}(\bm{x}, \bm{c}) = 0$ where $\bm{x}$ is a vector containing the spectral coefficients of $R$, $Z$, and $\lambda$, while $\bm{c}$ contains fixed parameters that define the equilibrium problem, such as the pressure and rotational transform profiles and the fixed boundary shape. 

Since it's original publication \citep{dudt2020desc}, DESC has undergone a major upgrade that involved porting it from MATLAB to Python in order to take advantage of the JAX library \citep{jax2018github} for automatic differentiation (AD) and just-in-time (JIT) compilation. Initially developed for machine learning applications, JAX provides an NumPy\citep{numpy2011,numpy2020}-like API for common mathematical operations and allows arbitrary functions to be differentiated using forward or reverse mode AD. This allows the calculation of exact derivatives of the objective function automatically, rather than having to code them by hand which is time-consuming and error prone, or using finite differences which are computationally expensive and can be inaccurate. It is also much more flexible as new objective functions can be added and optimized by defining only the forward pass, which is of great use in stellarator optimization where new objectives may be added in the future. JAX also allows for JIT compiling of code to both CPUs and GPUs which significantly speeds up calculation, approaching speeds of traditional compiled languages, avoiding one of the primary limitations of Python for scientific computing. Additionally, given that the vast majority of new supercomputers heavily leverage GPUs, and this trend is likely to continue, using JAX allows DESC to take full advantage of all the compute capability available, rather than being limited to CPU-only parallelization like many legacy codes. This allows a "best of both worlds" approach where the code is easy to use, maintain, adapt, and upgrade, while still being fast enough for production applications. 

Several aspects also make DESC the ideal code for implementing the perturbation method outlined in \autoref{sec:perturbations}:
\begin{enumerate}
    \item Using a pseudo-spectral discretization significantly reduces the number of independent variables, resulting in smaller Jacobian matrices.
    \item Using JAX allows fast and accurate computation of the required derivatives and Jacobian-vector products.
    \item Formulating the problem as a system of nonlinear equations and solving them in a least squares sense also effectively gives an extra order of derivative for free. 
\end{enumerate}

This last point can be seen by considering the least squares problem

\begin{equation}
\min_{\bm{x}} y(\bm{x}) \equiv \frac{1}{2} \bm{f}(\bm{x})^T \bm{f}(\bm{x})
\label{eq:least_squares_f}
\end{equation}

where $\bm{f}$ is a vector valued function, and $y$ is the sum of squares of the residuals of $\bm{f}$. The gradient of $y$ is given by 
\begin{equation}
    \frac{dy}{d \bm{x}} = \bm{f}^T \frac{d \bm{f}}{d \bm{x}}
\end{equation}
and its Hessian (matrix of second partial derivatives) is given by
\begin{equation}
    \frac{d^2y}{d \bm{x}^2} = \frac{d \bm{f}}{d \bm{x}}^T \frac{d \bm{f}}{d \bm{x}} + \bm{f}^T \frac{d^2 \bm{f}}{d \bm{x}^2}.
\end{equation}

Given that $\bm{f}$ is the function we are trying to minimize in the least squares sense, we can generally assume that the 2nd term in the Hessian is negligible compared to the first, the so-called "small residual approximation" \citep{nocedal2006numerical}. This gives an approximate Hessian $\frac{d^2 y}{d \bm{x}^2} \sim \frac{d \bm{f}}{d \bm{x}}^T \frac{d \bm{f}}{d \bm{x}}$, meaning that using only first derivative information about $\bm{f}$ gives us both first and second derivative information about $y$. In addition, the approximate Hessian this gives is always positive semi-definite, leading to a convex subproblem which is easy to solve.

\section{Perturbations} \label{sec:perturbations}

A general fixed-boundary equilibrium problem can be described by a set of parameters $\bm{c} = \{R_b, Z_b, p, \iota, \Psi \}$ where $R_b, Z_b$ are the $R,Z$ coordinates of the boundary surface, $p$ is the pressure profile, $\iota$ the rotational transform, and $\Psi$ the total toroidal flux through the torus. In many spectral equilibrium codes \citep{dudt2020desc,vmec1983}, the independent variables that define the equilibrium can be given by $\bm{x} = [R_{lmn},Z_{lmn},\lambda_{lmn}]$ where $R_{lmn}$, $Z_{lmn}$ and $\lambda_{lmn}$ are spectral coefficients of the flux surface positions and the poloidal stream function (indexed by $l$ in the radial direction, $m$ in the poloidal direction, and $n$ in the toroidal direction). The condition of MHD equilibrium can then be written as a (possibly vector valued) nonlinear algebraic equation involving the fixed parameters and independent variables, $\bm{f}(\bm{x},\bm{c})=0$. The function $f$ is the discretized form of the general MHD force balance $\bm{J} \times \bm{B} - \nabla p = 0$\citep{dudt2020desc}, or a condition on the gradient of the energy functional $W = \int_V (B^2/2\mu_0 + p) dV$ \citep{vmec1983}. 

Given a set of parameters $\bm{c}$ and a solution vector $\bm{x}$ which satisfy $\bm{f}(\bm{x},\bm{c}) = 0$, we wish to find how the equilibrium would change if the parameters $\bm{c}$ are perturbed to $\bm{c} + \Delta \bm{c}$. For instance, we may have a solution for a vacuum equilibrium and want to see how adding finite pressure changes it, or we may start with a 2D tokamak solution and add a 3D perturbation to the boundary shape to form a stellarator.

\renewcommand{\thefootnote}{\fnsymbol{footnote}}
We assume that the new equilibrium is given by $\bm{x} + \Delta \bm{x}$, and expand $f$ in a Taylor series \footnote{Some care must be taken here, as we are implicitly assuming that the function $f$ is at least $C^2$ continuous. The DESC code assumes the existence of nested flux surfaces so that there is an analytic mapping between real space and magnetic coordinates. In cases where the assumption of nested surfaces is violated, this mapping may not exist. However, we note that the force balance equations are still analytic functions of the dependent variables (it can be shown that they form a high order polynomial), though their physical meaning will be somewhat unclear in regions where nested surfaces don't exist} :

\begin{multline}
\label{eq:taylor}
\bm{f}(\bm{x}+\Delta \bm{x},\bm{c}+ \Delta \bm{c}) = \bm{f}(\bm{x},\bm{c}) + \frac{\partial \bm{f}}{\partial \bm{x}}\Delta \bm{x} + \frac{\partial \bm{f}}{\partial \bm{c}}\Delta \bm{c}  + \frac{1}{2}\frac{\partial^2 \bm{f}}{\partial \bm{x}^2}\Delta \bm{x}\Delta \bm{x} \\
+ \frac{1}{2} \frac{\partial^2 \bm{f}}{\partial \bm{c}^2}\Delta \bm{c}\Delta \bm{c} +
 \frac{\partial^2 \bm{f}}{\partial \bm{x}\partial \bm{c}}\Delta \bm{x}\Delta \bm{c} + O(\Delta x^3) .
\end{multline}

We wish to solve this equation for $\Delta \bm{x}$ such that $\bm{f}(\bm{x}+\Delta \bm{x},\bm{c}+ \Delta \bm{c}) = 0$. At first order this is a straightforward algebraic equation, but at higher orders it becomes a tensor polynomial equation which can be difficult or impossible to solve efficiently (for example, if  $\bm{f}$ is a vector valued function, just storing the 2nd derivative tensor in memory could require upwards of 100 GB). Instead of seeking a direct solution, we can try a perturbative approach, were we introduce an arbitrary small parameter $\epsilon$ and further expand $\Delta \bm{x}$ and $\Delta \bm{c}$ in a perturbation series in powers of $\epsilon$ (we assume that $\Delta \bm{c}$ is known a-priori and so only a first order term is required).
\begin{eqnarray}
\label{eq:expansion}
\Delta \bm{x} &=& \epsilon \bm{x}_1 + \epsilon^2 \bm{x}_2 + ...\\
\Delta \bm{c} &=& \epsilon \bm{c}_1. 
\end{eqnarray}

Plugging this into \autoref{eq:taylor} (and setting $\bm{f}(\bm{x},\bm{c}) = \bm{f}(\bm{x}+\Delta \bm{x}, \bm{c} + \Delta \bm{c})=0$) we get:
\begin{multline}
\label{eq:tayloreps}
0 = \frac{\partial \bm{f}}{\partial \bm{x}}(\epsilon \bm{x}_1 + \epsilon^2 \bm{x}_2) + \frac{\partial \bm{f}}{\partial \bm{c}}\epsilon \bm{c}_1  + \frac{1}{2} \frac{\partial^2 \bm{f}}{\partial \bm{x}^2}(\epsilon \bm{x}_1 + \epsilon^2 \bm{x}_2)(\epsilon \bm{x}_1 + \epsilon^2 \bm{x}_2) \\
+ \frac{1}{2} \frac{\partial^2 \bm{f}}{\partial \bm{c}^2}\epsilon \bm{c}_1 \epsilon \bm{c}_1  + \frac{\partial^2 \bm{f}}{\partial \bm{x}\partial \bm{c}}(\epsilon \bm{x}_1 + \epsilon^2 \bm{x}_2) \epsilon \bm{c}_1 + O(\Delta x^3) .
\end{multline}%

We can then collect powers of $\epsilon$ and set each order of $\epsilon$ to zero in turn. The first order equation gives:

\begin{eqnarray}
{0} &=&\frac{\partial \bm{f}}{\partial \bm{x}}\epsilon \bm{x}_1 + \frac{\partial \bm{f}}{\partial \bm{c}}\epsilon \bm{c}_1 \\
x_1 &=& -\Bigg(\frac{\partial \bm{f}}{\partial \bm{x}}\Bigg)^{-1} \Bigg( \frac{\partial \bm{f}}{\partial \bm{c}} c_1  \Bigg).
\end{eqnarray}

The second order term gives:

\begin{eqnarray}
{0} &=& \frac{\partial \bm{f}}{\partial \bm{x}}\epsilon^2 \bm{x}_2 + \frac{1}{2}\frac{\partial^2 \bm{f}}{\partial \bm{x}^2}\epsilon^2 \bm{x}_1 \bm{x}_1 + \frac{1}{2}\frac{\partial^2 \bm{f}}{\partial \bm{c}^2}\epsilon^2 \bm{c}_1 \bm{c}_1  + \frac{\partial^2 \bm{f}}{\partial \bm{x}\partial \bm{c}}\epsilon^2 \bm{x}_1 \bm{c}_1  \\
\bm{x}_2 &=& -\Bigg(\frac{\partial \bm{f}}{\partial \bm{x}}\Bigg)^{-1} \Bigg( \frac{1}{2}\frac{\partial^2 \bm{f}}{\partial \bm{x}^2} \bm{x}_1 \bm{x}_1 + \frac{1}{2}\frac{\partial^2 \bm{f}}{\partial \bm{c}^2} \bm{c}_1 \bm{c}_1  + \frac{\partial^2 \bm{f}}{\partial \bm{x}\partial \bm{c}} \bm{x}_1 \bm{c}_1  \Bigg) .
\end{eqnarray}

In general, the second derivative terms will be large, dense, rank 3 tensors (recall that $\bm$ can be a vector valued function), which may be extremely expensive to compute and may not even fit in memory for high resolution cases. However, it is important to note that the full second derivatives are never needed (or indeed full first derivatives apart from the Jacobian matrix $\partial \bm{f}/\partial \bm{x}$). All that is needed are directional derivatives or Jacobian vector products. In the DESC code (\autoref{sec:code_improvements}) these are calculated using forward mode automatic differentiation at a cost a few times greater than the cost of a single evaluation of $\bm{f}$. These terms could also be approximated using finite differences at a similar cost. This means that at each order the most expensive operation is solving a linear system of the form 
\begin{equation}
\bm{J} \bm{x}_i = \bm{b}
\end{equation}
where $\bm{J} \equiv \partial \bm{f} / \partial \bm{x}$ is the Jacobian matrix and is the same at each order, so only needs to be decomposed once and $i$ is the order of the term in the perturbation series. Because of these computation and memory savings, the method has been extended to 3rd order in the DESC code (see \autoref{sec:code_improvements}), though in most situations 1st or 2nd order perturbations gives acceptable results (see \autoref{fig:dshape}).

It is instructive to note that the first order term in the perturbation series for $\Delta \bm{x}$ is effectively the Newton step, while the second order term is commonly referred to as the Halley step in optimization literature \citep{gander1985halley,gundersen2010}. While Newton's method converges quadratically \citep{nocedal2006numerical} when started sufficiently close to a solution, if started far away it may diverge and so in practice the method must be globalized using either a line search or a trust region framework. Although we assume that we begin near or at an equilibrium so that $\bm{f}(\bm{x},\bm{c}) = 0$, depending on the size of the perturbation the solution landscape may change such that this is no longer the case, and we find in practice in several cases the unconstrained Newton step causes the solution to diverge. There is also an additional constraint inherent in the approach taken that the second order term should be smaller than the first order term by a factor $\epsilon << 1$.

This suggests that a natural way to enforce these conditions and ensure reasonable convergence properties is to use a trust region method. When expanding the objective function in a Taylor series we are effectively approximating it by a linear or quadratic function. Instead of seeking the global minimizer to these model functions, the trust region approach instead recognizes that the Taylor approximation is only valid in some small neighborhood and restricts the step size accordingly. Mathematically, instead of finding the exact solution $\bm{x}_i^*$ to the linear system 
\begin{equation}
    \bm{J} \bm{x}_i^* = \bm{b}.
\end{equation}
We instead seek a solution to the following optimization problem:
\begin{equation}
\min_{\bm{x}_i} ||\bm{J} \bm{x}_i - \bm{b}||^2 \quad \quad s.t. \quad ||\bm{x}_i|| \le r
\end{equation}
where $r$ is the radius of the trust region. This subproblem has two possible solutions \citep{nocedal2006numerical}: either the true solution $\bm{x}_i^*$ lies within the trust region, or else that there is a scalar $\alpha > 0$ such that 
\begin{equation}
    (\bm{J} + \alpha I)\bm{x}_i = \bm{b}, \quad \quad ||\bm{x}_i||=r.
\end{equation}
The constrained value for $\bm{x}_i$ can be efficiently found using a 1D root finding method in $\alpha$, in the DESC code we use a bracketed Newton method. The cost of solving this subproblem is roughly the cost of one singular value decomposition of the Jacobian $\bm{J}$, or two to three Cholesky factorizations of the approximate Hessian matrix $\bm{B} \equiv \bm{J}^T \bm{J}$. 

Selection of the trust region radius $r$ is traditionally done adaptively within an optimization loop, by comparing the actual reduction in the residual at each step with that predicted by the model function. In a perturbation, only a single "step" is taken, and we have found empirically that setting the trust region at a small fraction of $||\bm{x}||$ generally gives good results, typically $r=0.1 ||\bm{x}||$. For second and higher order perturbations, such a large trust region can result lead to inaccurate results. This can be understood by considering the expansion made in \autoref{eq:expansion}, where it was implicitly assumed that the 2nd order correction was a factor of $\epsilon$ smaller than the first order term. Using a large trust region for the first order term ($r=0.1 ||\bm{x}||$) and a smaller trust region for the higher order terms correctly enforces this assumed scaling. In practice we obtained good performance using $r_i=0.1 ||\bm{x}_{i-1}||$, so the second order term is an order of magnitude less than the first order term and so on for higher orders.

\section{Continuation Method} \label{sec:continuation}

Many MHD equilibrium codes \citep{vmec1983} \citep{spec2011} use a multigrid approach, where an initial guess is specified on a coarse grid, and the error is minimized on that grid before being interpolated to a finer grid and re-solved, continuing until the finest resolution level has been reached. This is done both to speed computation by doing more calculations on coarser grids, and also to make it more robust to poor initial guesses and avoid additional saddle points that can appear in high dimensional spaces \citep{dauphin2014}. DESC offers this as well, allowing the resolution to be varied in the radial, poloidal, and toroidal directions independently.

In addition to the standard multigrid approach, DESC has also implemented a new continuation method using the perturbation techniques described in \autoref{sec:perturbations}. In general, continuation methods seek to find how the solution to an equation varies as parameters are changed. Using the notation of \autoref{sec:perturbations} we can view $\bm{x}$ to be an implicit function of $\bm{c}$, related by the constraint equation $\bm{f}(\bm{x},\bm{c})=0$, and try to find the map that relates $\bm{x}$ and $\bm{c}$ along lines of $\bm{f}(\bm{x},\bm{c})=0$.

In this work, we note two properties that make continuation methods especially relevant: 
\begin{enumerate}
    \item 2D (axisymmetric) equilibria are much easier to compute than 3D equilibria, due to the guaranteed existence of flux surfaces in axisymmetric equilibria, and the significant reduction in the number of variables needed to represent the solution.
    \item Vacuum (vanishing beta) equilibria are easier to compute than finite pressure equilibria, due to the lack of Shafranov shift.
\end{enumerate}

We can take advantage of these properties to compute complicated stellarator equilibria by first solving an "easy" problem, and then introduce a continuation parameter that transforms our initial solution into the solution to a "hard" problem.

To do this, we introduce 2 scalar continuation parameters: $\eta_b$, a multiplier on all of the non-axisymmetric boundary modes, and $\eta_p$, a scaling factor for the pressure profile. Setting $\eta_b=0$ would give a 2D axisymmetric boundary that is "close" to the desired 3D equilibrium, while setting $\eta_p=0$ would give the zero pressure equilibrium with the desired boundary shape. By varying these two parameters, we find a "family" of solutions that are connected continuously.

To vary the parameters, we apply a sequence of perturbation steps as outlined in \autoref{sec:perturbations}. With each perturbation we take a small step in the desired direction in parameter space (eg, varying $\mathbf{c} = \{\eta_b, \eta_p\}$). Depending on the step size $\Delta \mathbf{c}$ it may be necessary to refine the solution with a small number (2-5) of Newton iterations of \autoref{eq:least_squares_f} to ensure a good equilibrium is reached. We then re-linearize about the new position and perform the next step, until the final desired parameters are reached.

A standard method for solving continuation problems, known as "natural parameter continuation" would be analogous to a 0th order perturbation followed by several Newton iterations of \autoref{eq:least_squares_f}, while a first order perturbation would be similar to Gauss-Newton continuation. The higher order perturbations discussed above do not seem to have been explored in the more general continuation method literature, but can be seen as a version of Halley's method applied to the combined system $\{\mathbf{f}(\mathbf{x},\mathbf{c})=0, \mathbf{c} - \mathbf{c}_{desired}=0\}$.

In most traditional continuation methods, the step size in the continuation parameter is determined adaptively based on local error estimates and the ratio of predicted to achieved error reduction. In practice we have found that this is often not necessary, and fixing the step sizes a-priori provides sufficiently accurate results. In most cases going from a zero pressure equilibrium to a moderate beta of $\sim 3\%$ can be done in a single step, and boundary perturbations in anywhere from 1-4 steps, depending on the desired accuracy and complexity of the boundary. This does require the user to specify the desired perturbation steps explicitly, though a future upgrade to the code is planned to allow adaptive perturbations where the user need only supply an initial guess and the final desired parameters.

\begin{figure*}
            \centering
	        \includegraphics[width=6in]{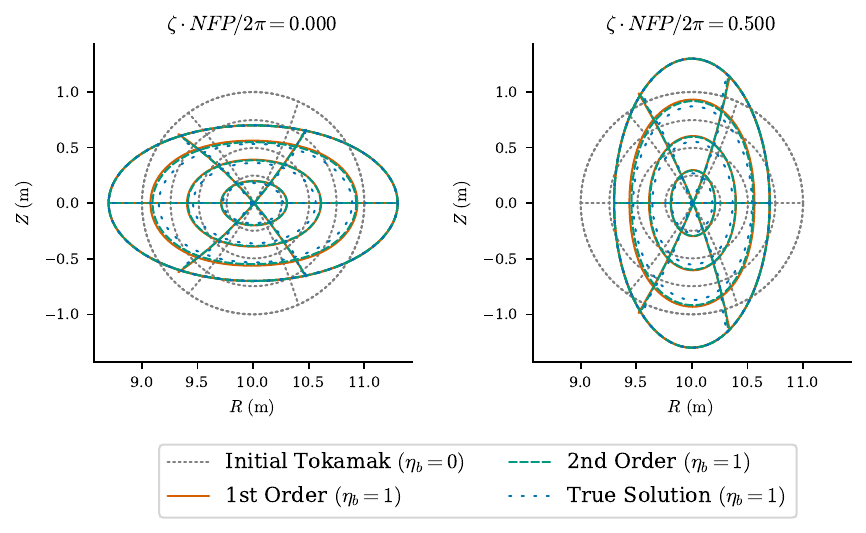}
	        \caption{Comparison of perturbed tokamak equilibrium vs true solution for a heliotron like stellarator. The initial tokamak (grey) is axisymmetric. By applying a perturbation (1st order in red, 2nd order in green) to the non-axisymmetric boundary modes, we obtain an approximation to the true 3D solution (blue), obtained by solving from the start with the full 3D boundary shape. The first order perturbation captures the majority of the differences between the initial and true solutions. The second order effects bring the perturbed solution even closer to the true one.}
            \label{fig:heliotron_bdry}
        \end{figure*}

As a first example (\autoref{fig:heliotron_bdry}), we solve for a zero pressure heliotron by first solving for a simple circular tokamak, then applying a 3D perturbation to the boundary. After the perturbation, a small number of regular Newton iterations of \autoref{eq:least_squares_f} are applied to ensure convergence. 

\begin{table}
\begin{tabular}{c|c}
    Perturbation order & Mean flux surface error ($\frac{1}{V} \int_V ||\mathbf{r}_{true} - \mathbf{r}_{perturbed}|| d^3\mathbf{r}$) \\
    \hline
    Initial tokamak &  0.100 m\\
    1st order &  0.031 m\\
    2nd order &  0.026 m\\
\end{tabular}
\caption{Mean flux surface position error vs perturbation order for boundary perturbation from a circular tokamak to a heliotron}
\label{table:heliotron_error}
\end{table}

Similarly, when computing finite pressure equilibria it can be difficult to estimate the Shafranov shift a-priori for generating a good initial guess. DESC avoids this by first computing a zero pressure solution, for which a good initial guess can generally be found by simply scaling the boundary surface. The finite pressure is then added back in as a perturbation, which then often only requires a small number of further iterations to reach convergence as shown in \autoref{fig:dshape}. In these and the following examples, the rotational transform profile is held fixed during the perturbations, resulting in a change in the toroidal current as the pressure and boundary are varied. In this example, the toroidal current before and after (not shown) is largely similar, with a small amount of additional current required to support the increased pressure.

\begin{figure}
	\includegraphics[width=3in]{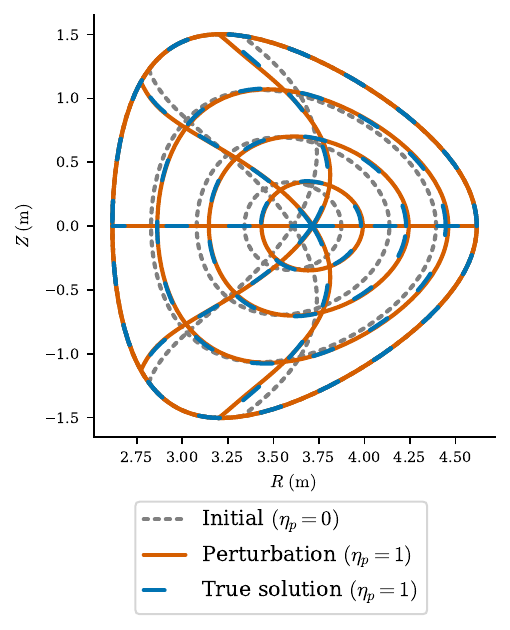}
	\centering
	\caption{Comparison of perturbed equilibrium vs true solution for a change in pressure from $\beta=0\%$ to $\beta=3\%$. The first order perturbation (red) captures the majority of the differences between the initial (grey) and true (blue) solutions. Second and third order effects (not shown for clarity) are visually very similar but significantly reduce the resulting force balance error as shown in \autoref{table:dshape_error}.}
	\label{fig:dshape}

\end{figure}

\begin{table}
\begin{tabular}{c|c}
    Perturbation order & Normalized force error ($|F| / |\nabla p|)$ \\
    \hline
    1st order & $87.6 \%$ \\
    2nd order & $29.6 \%$ \\
    3rd order & $9.7 \%$ \\
\end{tabular}
\caption{Normalized force balance error vs perturbation order for a pressure increase from $\beta=0\%$ to $\beta=3\%$.}
\label{table:dshape_error}
\end{table}

While in general the two parameters could be varied in any order, or simultaneously when solving for a 3D finite-beta equilibrium, we have found that first varying $\eta_p$, followed by $\eta_b$ to be more efficient. Varying $\eta_p$ while holding $\eta_b$ fixed at $0$ allows the pressure perturbations to be done on a 2D axisymmetric configuration, which reduces the computational cost and ensures the existence of good flux surfaces. After reaching a high resolution finite-beta axisymmetric equilibrium, 3D modes are added to the basis functions and the boundary is perturbed to give the final desired 3D finite-beta equilibrium. 

This procedure is demonstrated in \autoref{fig:w7x_bdry_scan}, where the initial solution is an axisymmetric zero pressure tokamak. The pressure is then increased, as demonstrated by the Shafranov shift, followed by 3D deformation of the boundary shape to arrive at a W7X like configuration. After each large perturbation, a small number of regular Newton iterations of \autoref{eq:least_squares_f} are performed to ensure that an equilibrium has been reached (recall that in the derivation of the perturbations, it was assumed that the initial state before perturbing was an equilibrium). This method was also used to compute the solutions shown in Part I \citep{panici2022desc}, where a detailed analysis of the force error is given.

\begin{figure}
	\includegraphics[width=3.5in]{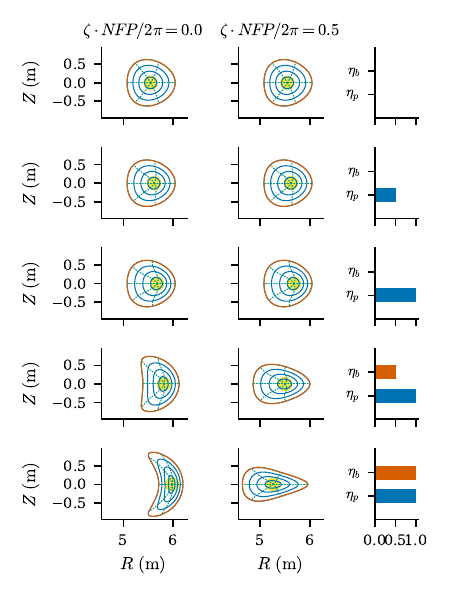}
	\centering
	\caption{Continuation solution for W7X like equilibrium. Starting from a zero pressure 2D equilibrium, the pressure is increased to $\beta \sim 3\%$ in a series of two steps. Then a 3D perturbation is applied to the boundary, broken up into 4 steps (2 shown) to arrive at the final finite $\beta$ 3D solution.}
	\label{fig:w7x_bdry_scan}

\end{figure}

\begin{figure}
	\includegraphics[width=3.5in]{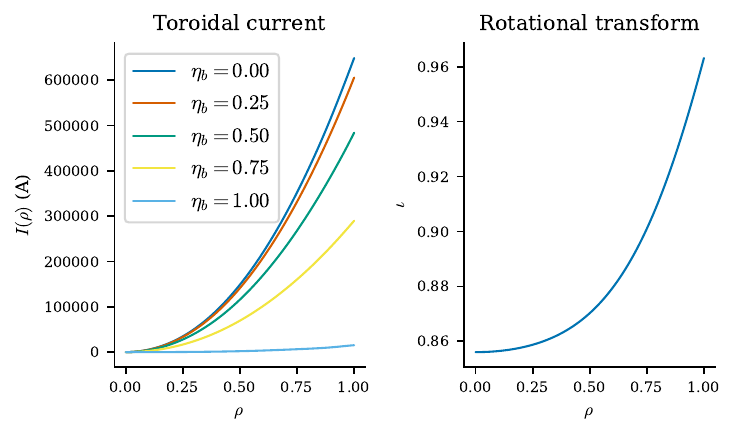}
	\centering
	\caption{Toroidal current and rotational transform for W7X like equilibrium with varying boundary ratios. When $\eta_b=0$, corresponding to an axisymmetric solution, a large toroidal current is required to generate the rotational transform (which is held fixed as the boundary is varied). Increasing the 3D shaping allows more of the rotational transform to come from external fields, reducing the plasma current. Note the current in the final step is still nonzero, as the rotational transform being held fixed was chosen for illustration, and does not correspond to the vacuum rotational transform on W7-X.}
	\label{fig:w7x_current}

\end{figure}

Further insight can be gained by looking at the toroidal current profile of the W7X like equilibrium as the boundary ratio is varied in \autoref{fig:w7x_current}. At each step we keep the rotational transform profile fixed, and so as expected the axisymmetric case ($\eta_b=0$) requires a large toroidal current to generate the poloidal field. As we increase the 3D shaping, more of the rotational transform is generated by axis torsion and boundary shaping, reducing the required plasma current to near zero.

An important feature to note about the aforementioned continuation method is that the solution at each step (after any necessary Newton iterations of \autoref{eq:least_squares_f} to refine the solution) is in fact the "exact" solution to the equilibrium problem with perturbed parameters, and is hence still a valid equilibrium that satisfies MHD force balance and may have desirable physics properties worth studying that may be absent in the final solution. By applying different perturbations and varying different parameters, one can explore whole families of solutions that are "nearby" to a starting equilibrium.

\section{Other Applications} \label{sec:applications}

\subsection{Parameter Scans} \label{sub:scans}

Another important feature of continuation methods is revealing how solutions change as parameters of the problem are varied. In the previous section we used this to find single equilibria, but the method can also be used to explore families of equilibria related by a parameter or group of parameters. In this, we find equilibria "nearby" to an equilibrium already found, such as examining the same boundary shape at different values of $\beta$ or boundary perturbations of varying magnitude and shape such as resonant magnetic perturbations (RMP) in a tokamak. Traditionally this requires re-solving the equilibrium for each new value of the parameter. An alternative approach is to apply a perturbation to an initial equilibrium and find the corresponding changes in the flux surfaces. After applying the first perturbation, we re-linearize about the new state and perturb again, and so step through different equilibria for the cost of a single linear system solve at each step. This represents a significant computational savings compared to solving the full nonlinear problem for each value of the parameter.

\begin{figure}
	\includegraphics[width=3in]{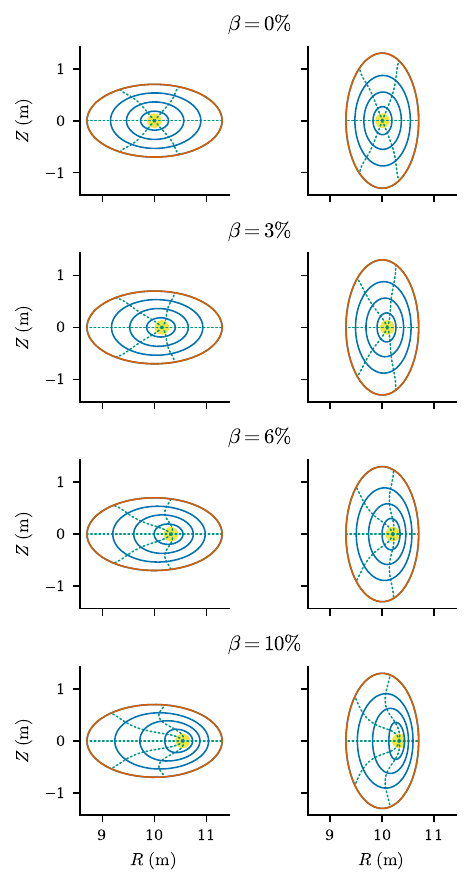}
	\centering
	\caption{Flux surfaces on the $\zeta=0$ and $\zeta=\pi$ planes for a helical stellarator. The initial equilibrium had $\beta=0\%$, and a series of 2nd order perturbations were applied to increase the pressure rather than solving a new equilibrium from scratch each time.}
	\label{fig:pres_scan}
\end{figure}

\autoref{fig:pres_scan} shows the flux surfaces on the $\zeta=0$ and $\zeta=\pi$ plane for a helical stellarator for the same pressure profile scaled to different values of $\beta$. The initial solution was at $\beta = 0\%$, and 2nd order perturbations were applied sequentially to step the pressure up to $\beta=10\%$. As in previous examples, the rotational transform is held fixed as the pressure is increased, leading to an increase in the toroidal current. The flux surfaces are indistinguishable from those obtained by solving from scratch at each value of $\beta$, and the computational cost is significantly reduced, as shown in \autoref{tab:heliotron_scan_times}. Performing such a scan requires only a few lines of code, as shown in \autoref{lst:perturbbeta}.

\begin{lstlisting}[language=Python, 
caption=Code example for $\beta$ scan, label={lst:perturbbeta}]
import numpy as np
import desc.io
from desc.equilibrium import EquilibriaFamily
eq0 = desc.io.load("heliotron_vacuum_solution.h5")[-1]
eqf = EquilibriaFamily(eq0)
# this corresponds to a change in beta of  ~1%
dp = np.array([ 1800., 0., -3600., 0., 1800.]) 
for i in range(10):
    eqf.append(eqf[-1].perturb(dp=dp, order=1))
    # polish off solution 
    eqf[-1].solve(maxiter=5) 
\end{lstlisting}

\begin{table}
    \centering
    \begin{tabular}{c|c|c}
        $\beta$ & Time (s) &  Time (s)  \\
        & without perturbations & with perturbations \\
        \hline
        1 & 164 & 78.0 \\
        2 & 174 & 16.3 s \\
        3 & 164 & 14.4 s \\
        4 & 122 & 12.6 s \\
        5 & 127 & 17.8 s \\
        6 & 122 & 19.1 s \\
        7 & 121 & 15.2 s \\
        8 & 134 & 13.6 s \\
        9 & 140 & 16.0 s \\
        10 & 138 & 18.3 s \\
        \hline
        Total & 23.4 min & 3.68 min \\
    \end{tabular}
    \caption{Computation times for different values of $\beta$ in the pressure scan shown in \autoref{fig:pres_scan}, both with and without perturbations, starting from a zero pressure solution ($\beta=0$, not shown). The first perturbation takes longer as the time includes JIT compilation for the specific problem being solved, resulting in significantly faster times for the subsequent perturbation steps which reuse the same compiled code. The "without perturbations" method is already able to take advantage of compiled code from the initial solve and does not see any significant speedup at subsequent iterations. All computation done on an AMD Ryzen 7 PRO 4750U with 32 GB memory.}
    \label{tab:heliotron_scan_times}
\end{table}

\begin{figure}
	\includegraphics[width=3in]{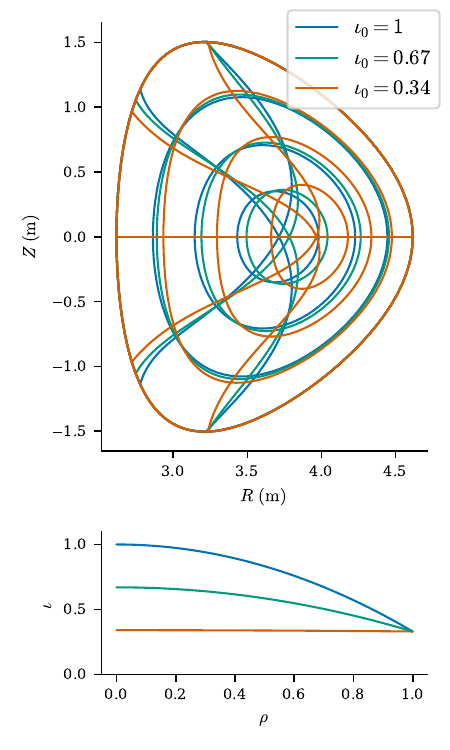}
	\centering
	\caption{\textbf{Top}: Flux surfaces on the $\zeta=0$ plane for a D-shaped tokamak with varying axis rotational transform. The initial equilibrium (blue) corresponds to $\iota_0=1$, and perturbations were used to reduce the axis rotational transform without recomputing the entire equilibrium solution. \textbf{Bottom}: Rotational transform profiles corresponding to flux surfaces shown above.}
	\label{fig:iota_scan}
\end{figure}

As another example (\autoref{fig:iota_scan}), a D-shaped tokamak was used as the starting point for variations in the rotational transform profile. The initial equilibrium had a rotational transform on axis of $\iota_0=1$, and perturbations were applied to reduce this down to $\iota_0=0.34$. As in the previous examples, the prescribed profiles at each step are pressure and rotational transform. As the rotational transform is decreased, we see increased Shafranov shift and deformation of the flux surfaces as the reduced poloidal field struggles to contain the pressure ($\beta \sim 3\%$ in this example).

As in the pressure scan example, the solution obtained by perturbation is indistinguishable by eye from the solution obtained by solving the full equilibrium problem, and significantly faster. Once an initial equilibrium is solved, parameter scans can be performed an order of magnitude faster using perturbations than other approaches which require a full solution at each step. The code to perform such a scan is shown in \autoref{lst:perturbiota}.

\begin{lstlisting}[language=Python, caption=Code example for perturbing iota profile, label={lst:perturbiota}]
import numpy as np
import desc.io
eq0 = desc.io.load("DSHAPE.h5")[-1]
# perturbing a 2nd order polynomial basis for iota
di = np.array([-0.33, 0, 0.33]) 
eq1 = eq0.perturb(di=di, order=2)
eq1.solve(maxiter=5) 
eq2 = eq1.perturb(di=di, order=2)
eq2.solve(maxiter=5)
\end{lstlisting}

\subsection{Optimization} \label{sub:optimization}

The perturbation techniques described previously can also be adapted for optimization, where instead of choosing a particular change in the parameters $\Delta \bm{c}$ we instead let $\Delta \bm{c}$ be a free parameter that is chosen to minimize a cost function $\bm{g}(\bm{x},\bm{c})$, such as quasisymmetry error \citep{boozer1983,helander2014}, coil complexity \citep{zhu2018focus,zhu2018hessian,mcgreivy2020}, or fast particle confinement \citep{nemov2008,velasco2021}:

\begin{equation}
    \Delta \bm{c}^* = \argmin_{\Delta \bm{c}} \bm{g}(\bm{x}+\Delta \bm{x}, \bm{c} + \Delta \bm{c}) \quad s.t. \quad \bm{f}(\bm{x}+\Delta \bm{x}, \bm{c}+\Delta \bm{c}) = 0
\end{equation}
Where as before $\Delta \bm{x}$ is an implicit function of $\Delta \bm{c}$.

This finds the step in parameter space that most decreases the cost function $\bm{g}$ while maintaining approximate force balance. After applying this perturbation step, a small number of Newton iterations of \autoref{eq:least_squares_f} are used to re-converge to the correct equilibrium, without the need for a full "cold start" equilibrium solve. This single "warm start" equilibrium solve in DESC can be contrasted with the method of STELLOPT \cite{stellopt} or SIMSOPT \cite{simsopt2021} where at each optimization step a series of $N$ cold start equilibrium solves must be performed to find the descent direction, where $N$ is the number of variables being optimized.

In the optimization literature, this is part of a more general class of methods for constrained optimization. Like the perturbations described in \autoref{sec:perturbations}, this method can also be extended to higher order. This extension and further details on this method of optimization and applications to quasisymmetry are given in Part III \citep{dudt2022desc}.

\section{Conclusions}
\label{sec:conclusion}
We have demonstrated a new technique for computing stellarator equilibria, and for exploring how those equilibria change as parameters are varied. The methods are computationally efficient, and offer significant speedups compared to existing techniques, and in many cases offer possibilities that have not existed before. An important future application of these methods is in exploring the connections between different classes of equilibria, such as how tokamaks bifurcate into stellarators, and how different classes of quasisymmetric stellarators may be related.

\section*{Funding}
This work was supported by the U.S. Department of Energy under contract numbers DE-AC02-09CH11466, DE- SC0022005 and Field Work Proposal No. 1019. The United States Government retains a non-exclusive, paid-up, irrevocable, world-wide license to publish or reproduce the published form of this manuscript, or allow others to do so, for United States Government purposes.

\section*{Declaration of interests}
The authors report no conflict of interest.

\section*{Data availability statement}
The source code to generate the results and plots in this study are openly available in DESC at https://github.com/PlasmaControl/DESC or http://doi.org/10.5281/zenodo.4876504

\section*{Author ORCID}
R. Conlin, https://orcid.org/0000-0001-8366-2111; D. Dudt, https://orcid.org/0000-0002-4557-3529; D. Panici, https://orcid.org/0000-0003-0736-4360; E. Kolemen, https://orcid.org/0000-0003-4212-3247

\bibliographystyle{jpp}

\bibliography{bibliography}

\end{document}